# BROKEN - SYMMETRY - ADAPTED GREEN FUNCTION THEORY OF CONDENSED MATTER SYSTEMS:
## TOWARDS A VECTOR SPIN - DENSITY - FUNCTIONAL THEORY

A. K. Rajagopal
Naval Research Laboratory, Washington D. C. 20375-5320

Mogus Mochena
Department of Physics, Florida A & M University, Tallahassee, Florida 32307

## ABSTRACT

The group theory framework developed by Fukutome for a systematic analysis of the various broken symmetry types of Hartree-Fock solutions exhibiting spin structures is here extended to the general many-body context using spinor-Green function formalism for describing magnetic systems. Consequences of this theory are discussed for examining the magnetism of itinerant electrons in nanometric systems of current interest as well as bulk systems where a vector spin-density form is required, by specializing our work to spin-density-functional formalism. We also formulate the linear response theory for such a system and compare and contrast them with the recent results obtained for localized electron systems. The various phenomenological treatments of itinerant magnetic systems are here unified in this group-theoretical description.

## I. INTRODUCTION:

Irreducible representations of the symmetry group of a system which classify the eigenstates of its Hamiltonian have been discussed before [1]. However, many important physical phenomena such as superconductivity and many types of magnetic states of condensed matter display lower symmetry characteristics, and are thus broken symmetry states of the Hamiltonian [1]. These phenomena have been theoretically understood in terms of approximate solutions which break the full symmetry of the Schrödinger equation of the many - body system. For example, the celebrated Hartree - Fock (- Bogoliubov) self-consistent field method has been used in this way. A clear description of this approach may be found in [1] where an enumeration of the various commonly found basic symmetries is also given. Fukutome [2] and his coworkers, based only on symmetry arguments, systematized the search for possible novel states of the system, which in the past depended on intuitive suggestions ( viz. Bardeen, Overhauser, and Landau, for example). These authors developed a complete group-theoretical classification and characterization of all the possible magnetic symmetry structures arising from only the underlying symmetry group  consisting of spin rotations (S), time reversal (T), and spatial symmetry group (P) and its subgroups, within the Hartree - Fock approximation. Fukutome found that there are only eight subgroups of the symmetry group $S \times T$, consisting of spin rotations and time reversal. The Hamiltonian of the well-known electron gas which is an important ingredient in the density-functional method, is invariant under $S \times T$. It is common knowledge that the electron gas exhibits several states of magnetic order when treated in HF and related mean field approximation [3]. The eight different structures found correspond to those belonging to the eight subgroups. When the spatial symmetry group is combined with these, one gets many more varieties of structures. This theory has been more recently extended to include mean - field type analysis of Hubbard - type models of one- and two - dimensional systems of great current interest [4]. In recent years, spin - density - functional method [5] has become the method of choice replacing the Hartree - Fock (HF) method both because it includes the important correlation effects not included in the HF - scheme in describing the theory of magnetic systems and because of the present much improved computing techniques. There has been a suggestion to incorporate the Fukutome classification to the spin - density - functional formalism [6]. The Green function method for describing the many - body systems is more general than the above schemes and the broken - symmetry solutions can be incorporated in this framework as noted already in [1]. A generalized version of the density - functional method which includes spin as well has been reformulated in the Green function language [7]. The various broken symmetry states of the HF scheme will be more generally expressed here in terms of the symmetry of the Green function under the appropriate subgroups for which we give the nomenclature "broken-symmetry-adapted Green functions." The purpose of this paper is two fold; one, to incorporate the Fukutome classification into the Green function framework and, two, to focus attention on magnetic systems in general, by applying this formalism to the spin-density-functional method with an emphasis on the currently interesting findings of magnetic states in nanometric systems [8]. By this extension, we also provide a more complete vector - spin - density - functional (VSDF) theory of magnetic systems which goes beyond the currently used schemes for handling noncollinear magnetic materials [9, 10]. The recent theoretical work on the magnetism of free atomic clusters exhibiting interesting geometric structures accompanying very large magnetic moments [11] provides another class of problems in magnetism requiring further investigation particularly if the suggestion that these clusters can be formed on semiconductor interfaces [12]. In this case the possibilities of variety of geometric structures of the cluster formation each with its own magnetic features in relation to the substrate would provide some challenge in determining the appropriate broken symmetry states of magnetism in such environment. The group theoretical enumeration would make this search systematic and orderly.

    In Section II, we first give a brief description of the magnetic system in terms of the spinor Green function [13] and the associated spinor self - energy operator [14]. We then develop the Fukutome classifications of magnetic systems associated with this spinor Green function and the related spinor self - energy (or

mass -) operator, which is more general than the effective self - consistent potential of the HF theory. In Section III, the linear response functions arising out of these Green functions [15] are examined to establish the stability aspects of the broken symmetry structures. This will be shown to be related to the recent work [16]. In Section IV, the implications of the above developments to the VSDF theory are spelled out. This is complementary to the work in [17]. In the final Section V, a summary and concluding remarks in relation to the proposed work on magnetism of nanostructure systems including magnetic atomic clusters are given.

## II. THE BROKEN SYMMETRY ADAPTED GREEN FUNCTIONS - THE FUKUTOME CLASSIFICATIONS IN MAGNETIC SYSTEMS

### (a) Hamiltonian:

The general spin polarized system without spin - orbit interaction (for simplicity of presentation at this juncture) is described by the standard system Hamiltonian

$$H = T_e + V_{ii} + V_{ie} + V_{ee}. \tag{1}$$

Here, $T_e$ is the kinetic energy operator of the electrons, $V_{ii}, V_{ie},$ and $V_{ee}$ are operators representing the Coulomb interactions between the ions (i) and the electrons (e). Introduce $\psi_s(\vec{r})$, the usual field operator annihilating an electron of spin s at position $\vec{r}$, and $\psi_s^\dagger(\vec{r})$, the corresponding creation operator. Then we have the definitions of electron density operator

$$\hat{n}(\vec{r}) = \hat{s}_o(\vec{r}) = \sum_s \psi_s^\dagger(\vec{r}) \psi_s(\vec{r}), \tag{2}$$

and the vector spin - density - operator

$$\vec{s}(\vec{r}) = \sum_{s,s'} \vec{\tau}_{ss'} \psi_{s'}^\dagger(\vec{r}) \psi_s(\vec{r}), \tag{3}$$

where the Pauli spin matrix vector is $\vec{\tau}$. It is useful to consider the system as being subjected to external, spin-dependent field described by

$$V_{ext} = \sum_{s,s'} \int d\vec{r}\, w_{ss'}(\vec{r}t) \psi_s^\dagger(\vec{r}) \psi_{s'}(\vec{r}). \tag{4}$$

In the following we split $w_{ss'}$ into a scalar part $w_n \equiv f^0$ which acts on the electron density given by eq.(2) and a traceless part $w_S$ expressed in terms of the Pauli matrices which acts on the spin density, eq.(3):

$$w_S(\vec{r}t) = \vec{\tau} \cdot \vec{f}(\vec{r}t). \tag{5}$$

We thus rewrite eq.(4) in a physically transparent form

$$V_{ext} = \int d\vec{r} \left[ \hat{n}(\vec{r}) w_n(\vec{r}t) + \vec{s}(\vec{r}) \cdot \vec{f}(\vec{r}t) \right]$$
$$= \sum_\alpha \int d\vec{r}\, \hat{s}_\alpha(\vec{r}) f^\alpha(\vec{r}t), \tag{6}$$

where we have introduced the 4-vector notation

$$\left(f^{\alpha}\right) \equiv \left(w_n, \vec{f}\right), \; \left(\tau_{\alpha}\right) \equiv \left(1, \vec{\tau}\right), \; and$$

$$\hat{s}_{\alpha}(\vec{r}) \equiv \sum_{s,s'} \tau_{\alpha,ss'} \psi_{s'}^{\dagger}(\vec{r}) \psi_s(\vec{r}). \tag{7}$$

In the above Greek indices run from 0 to 3. The Hamiltonian (1) is invariant under all spin rotations, S, and time-reversal, T.

**(b) Spinor-Green's Function and the Spinor-Self Enegy:**

The one-particle Green's function is written as a matrix in spin indices [13],

$$G_{ss'}(\vec{r}t; \vec{r}'t') = -i\left\langle T\left(\psi_s(\vec{r}t)\psi_{s'}^{\dagger}(\vec{r}'t')\right)\right\rangle. \tag{8}$$

The equation of motion satisfied by this spinor-Green function is often written in its most general form as

$$\left[i\frac{\partial}{\partial t_1} + \frac{\vec{\nabla}_1^2}{2m} - V_C(1) - w(1)\right]G(12) - \int d(3)\Sigma(13)G(32) = \delta(12). \tag{9}$$

Here the spin-independent classical Coulomb potential $V_C$ arising from the electron density $n(1) = \langle \hat{n}(1) \rangle = -iTrG(11^+)$ and the nuclei ($V_{ion}$) is given by

$$V_C(1) = \int (d2) v(12) n(2) + V_{ion}(1), \tag{10}$$

where the instantaneous Coulomb interaction between electrons is $v(12) = \frac{1}{|\vec{r}_1 - \vec{r}_2|}\delta(t_1 - t_2)$. Also, Tr stands for trace over spin indices only. Also $w(1) = w_n(1) + \vec{\tau}\cdot\vec{f}(1)$, is the spinor representation of the external field. We have used here 1 to stand for the space-time point $(\vec{r}_1, t_1)$ and the other notations are standard usage as in [13], for example. The last term in the left hand side of eq.(9) is the contribution due to inter-particle interaction in its most general form and we call "exchange-correlation spinor-self-energy" contribution, in anticipation of later application to functional formalism. The general expression for this spinor-self-energy matrix, $\Sigma$, is in general a functional of the spinor-Green function, G, and is given by

$$\Sigma(12) = i\int d(3)\int d(4) G(13)\Lambda_n(32;4)W(41) \tag{11}$$

with $\quad \Lambda_n(12;3) = -\frac{\delta G^{-1}(12)}{\delta V_{tot}(3)} \tag{12}$

is the screened charge-response vertex function, with $V_{tot} = V_C + w_n$. W is the Coulomb interaction $v(12)$, screened by the dielectric function, $\varepsilon$, both of which are spin scalars, defined by

$$\varepsilon^{-1}(12) = \frac{\delta V_{tot}(1)}{\delta w_n(2)}, \tag{13}$$

$$W(12) = \int d(3) \varepsilon^{-1}(13) v(32). \tag{14}$$

Finally we write the Dyson equation (9) in the familiar form

$$G^{-1}(12) = \left[ i\frac{\partial}{\partial t_1} + \frac{\vec{\nabla}_1^2}{2m} - V_C(1) - w(1) \right]\delta(12) - \Sigma(12) \tag{15}$$

$$\equiv G_0^{-1} - \Sigma.$$

In the usual HF approximation, $\Sigma$ is the familiar unscreened exchange self-energy if we do not include the screening by the dielectric function of the medium: $\Sigma(12) \approx \Sigma_{HF}(12) = iv(12)G_{HF}(12)$. This exhibits explicit dependence of the self energy on the Green function and hence the self-consistency feature of eq.(15). In [13], the spinor structure of G was used to construct a self-consistent solution to the general HF approximation in the Green function language, from which the general Overhauser's spiral spin-density-wave solution was deduced. It was then recognized as a "broken symmetry" solution of the HF equation because this solution breaks the spin-rotation and time-reversal symmetries of the interacting electron gas Hamiltonian! We should also point out that in constructing any approximation scheme for the self-energy $\Sigma$, it is useful to have certain conservation principles as well as variational character. Baym [14] set up such a conserving scheme by introducing a functional $\Phi[G]$ whose first variational derivative with respect to G yields the exact $\Sigma$ and constructed an expression for the grand potential

$$\Omega[G] = Tr\,tr\{\ln(-G)\} - Tr\,tr\Sigma G + \Phi[G] \tag{16}$$

which is stationary with respect to variations in G. Thus,

$$V_C + \Sigma = \frac{\delta \Phi}{\delta G}, \tag{17}$$

and the grand potential is constructed with $\Phi[G]$ as the building block such that it is stationary for variations of the spinor G that satisfies the exact matrix Dyson equation, eq.(15). Here tr stands for the same notation as in Baym [14], namely,

$$trAB = \int_0^{-i\beta} d1 \int_0^{-i\beta} d2\, A(12)B(21^+), \tag{16a}$$

$\beta$ is the inverse temperature. Also, as in Baym, the choice of the branch of the logarithm is such that the variation of the first term in eq.(16) is taken to be of the form $\delta\{tr\ln(-G)\} = -tr\{(\delta G^{-1})G\}$. This formulation of the many-body theory is known in the literature as the $\Phi$–derivable method and has recently been generalized to include the density functional formalism in [14a,b]. With this introduction of the one-particle spinor-Green function and the corresponding spinor-self-energy of the system under consideration, we are now in a position to generalize the Fukutome symmetry considerations of the HF solutions and subsequent classification of the broken symmetry solutions of a magnetic system.

Following [13], the most general forms of the spinor-Green function and the spinor-self energy function with no spin-symmetry considerations are expressed in terms of the Pauli matrices:

$$G(12) = \frac{1}{2}\{g_n(12) + \vec{\tau} \cdot \vec{g}_S(12)\}, \tag{18}$$

$$\Sigma(12) = \{\sigma_n(12) + \vec{\tau} \cdot \vec{\sigma}_S(12)\}, \tag{19}$$

We first note that the physical electron density and the physical vector spin density are respectively given by

$$n(1) = \langle \hat{n}(1) \rangle = -iTrG(11^+) = -ig_n(11^+). \tag{20}$$

and $\quad \vec{s}(1) = \langle \vec{s}(1) \rangle = -iTr\{\vec{\tau}G(11^+)\} = -i\vec{g}_S(11^+). \tag{21}$

The corresponding physical particle current vector and the physical spin current tensor respectively are given by

$$\vec{j}(1) = \langle \vec{j}(1) \rangle = -\frac{1}{2m} Tr\{(\vec{\nabla}_1 - \vec{\nabla}_{1'})G(11')\}\Big|_{1'=1^+}$$
$$= -\frac{1}{2m}\{(\vec{\nabla}_1 - \vec{\nabla}_{1'})g_n(11')\}\Big|_{1'=1^+}. \tag{22}$$

$$\overset{\text{t}}{j}_S(1) = \langle \overset{\text{t}}{j}_S(1) \rangle = -\frac{1}{2m} Tr\{\vec{\tau}(\vec{\nabla}_1 - \vec{\nabla}_{1'})G(11')\}\Big|_{1'=1^+}$$
$$= -\frac{1}{2m}\{(\vec{\nabla}_1 - \vec{\nabla}_{1'})\vec{g}_S(11')\}\Big|_{1'=1^+}. \tag{23}$$

Here we have expressed these quantities of interest in terms of the corresponding spin-scalar and spin-vector components of the full spinor-Green function, eq.(18). From eq.(9), we have

$$\left\{\frac{\partial}{\partial t_1} + \frac{1}{2mi}\vec{\nabla}_1 \cdot (\vec{\nabla}_1 - \vec{\nabla}_{1'})\right\}G^{\langle}(11')\Big|_{1'=1^+} -$$
$$-\frac{1}{2i}\{(\vec{\tau} \cdot \vec{f}(1))G^{\langle}(11^+) - G^{\langle}(11^+)(\vec{\tau} \cdot \vec{f}(1))\} = \tag{24}$$
$$= \frac{1}{i}\int d(2) \begin{bmatrix} \Sigma^{\rangle}(12)G^{\langle}(21^+) + G^{\langle}(12)\Sigma^{\rangle}(21^+) \\ -\Sigma^{\langle}(12)G^{\rangle}(21^+) - G^{\rangle}(12)\Sigma^{\langle}(21^+) \end{bmatrix}.$$

From this, using eqs. (18, 19), we deduce the following general continuity equations relating the densities and currents given in eqs.(20, 21, 22, 23):

$$\frac{\partial n(1)}{\partial t_1} + \vec{\nabla}_1 \cdot \vec{j}(1) = -\int d(2) \begin{Bmatrix} \sigma_n^{\rangle}(12)g_n^{\langle}(21^+) + \ldots \\ +\vec{\sigma}_S^{\rangle}(12) \cdot \vec{g}_S^{\langle}(21^+) + \ldots \end{Bmatrix} \tag{25}$$

$$\frac{\partial \vec{s}(1)}{\partial t_1} + \vec{\nabla}_1 \cdot \overset{\text{t}}{j}_S(1) - (\vec{f}(1) \times \vec{s}(1)) = -\int d(2) \begin{Bmatrix} \vec{\sigma}_S^{\rangle}(12)g_n^{\langle}(21^+) + \ldots \\ +i\vec{\sigma}_S^{\rangle}(12) \times \vec{g}_S^{\langle}(21^+) + \ldots \end{Bmatrix}$$
(26)

The dots in the above denote other terms arising from the indicated manipulations required to express eq.(24) in terms of the particle- and spin-densities. Eq.(25) is the usual continuity equation for the particle density while eq.(26) is the corresponding one for the vector spin-density. The third term in the left hand side is the torque term due to external field acting on the spin vector while the right hand side involves interaction contributions which are of two kinds. The first one is due to spin vector modified by the particle density while the second one is the cross product of the spin vector with the field due to interactions. The difference between the itinerant and the localized spin cases thus becomes clear. In the itinerant case, the divergence terms and the others on the right side contribute to give rise to spin wave dispersion while in the localized case, the divergence term is absent (no spin current) and only the cross product of spins contribute in the right hand side. We will return to this in the next section.

### (c) Fukutome's Classification of the Broken Symmetry Adapted Green functions and the Corresponding Self Energies

Consider the Dyson equation in the absence of external fields obtained by dropping w in eq. (15). It transforms under any unitary transformation, U, to the form

$$G'^{-1} = G_0'^{-1} - \Sigma', \text{ where } G_0'^{-1} = U\left[i\frac{\partial}{\partial t_1} + \frac{\overset{\pm}{\nabla}_1^2}{2m} - V_C(1)\right]\delta(12)U^\dagger, \tag{27}$$

$$G' = UGU^\dagger, \text{ and } \Sigma' = U\Sigma U^\dagger = \Sigma[G'].$$

The Hamitonian, H, given by eq.(1), leads to the self-energy as well as the first term in the right hand side of eq.(27). Now $G_0'^{-1} = G_0^{-1}$ because it is the noninteracting part; the self energy, on the other hand, reflects the contributions due to interactions, and is a therefore a functional of G, and has the form given in eq.(11). Two cases arise as with the HF theory [1]: (a) G'=G, i.e., G commutes with U and thus is invariant under the transformation, U, and $\Sigma' = \Sigma$. In this case, U represents the self-consistent symmetry of the corresponding $\Sigma$-scheme (compare, HF scheme [1]); (b) G is not invariant under U but leads to G' and $\Sigma'$ obeying the same form of the equation as G. Then U represents a **broken** symmetry. Fukutome [2] pointed out in the context of HF scheme that if $U_1$ belongs to a subgroup of the group of transformations U, then $G'_{HF}$ and the corresponding self-consistent $\Sigma'_{HF}$, form the broken symmetry adapted solutions! From this it is clear that such a scheme holds for general self energy that appears in the equation for the Green function as in eq.(15), which we now explore in detail in this work. We will now present this generalized version of the Fukutome [2] analysis and describe these broken symmetry possibilities.

We first consider only the group S of all spin rotations $\{I, U\}$, $I$ being the unit operator in the rotation group, and the group T of time reversal operation $\{I, T\}$ which is of order two, I is the unit operator and $T = i\tau_y C$, $C$ = *complex conjugation operator,* under which the Hamiltonian (1) is invariant. As in [2], the inclusion of spatial symmetry has to be dealt with individually depending on the type of spatial symmetry one wants to consider. The general consideration involving only S and T is sufficient for the present and indeed, as will be discussed in Sec.IV, the results obtained here form the basis for the vector-spin-density-functional theory of magnetic systems, where the interacting electron gas system provides the underlying functional for investigating the properties of many important inhomogeneous systems (see for example, [9, 10]). Since S commutes with T, the group of all S and T is the direct product group, $\mathcal{G} \equiv S \times T = \{\mathcal{I} \times \mathcal{I}, U, UT\}$. There are eight different subgroups of $\mathcal{G}$ which we will now enumerate. We note at once that $\mathcal{I} \times \mathcal{I}$, S, and T individually are three subgroups of G. The spin rotations A($\hat{e}$), about a fixed axis denoted by the unit vector $\hat{e}$, is a subgroup of S: A($\hat{e}$)=$\{U(\hat{e},\theta)\}$, where $\{U(\hat{e},\theta)\}$ is given by

$$U(\hat{e},\theta) = \exp[-i\theta/2(\vec{\tau}\cdot\hat{e})] = \cos\frac{\theta}{2} - i(\vec{\tau}\cdot\hat{e})\sin\frac{\theta}{2}. \tag{28}$$

The fifth subgroup M($\hat{e}'$) is of order two consisting of unit operator and the combined operation of $T$ with a spin rotation through an angle $\pi$ around an axis $\hat{e}'$: M($\hat{e}'$)=$\{I, TU(\hat{e}',\pi)\}$. There are two more subgroups that arise from the product of the elements A($\hat{e}$) and M($\hat{e}'$):

$$A(\hat{e}) \times M(\hat{e}') = \{U(\hat{e},\theta), U(\hat{e},\theta)U(\hat{e}',\pi)T\} \tag{29}$$

This gives rise to two groups, one, when the unit vectors are orthogonal, $\hat{e}\cdot\hat{e}' = 0$, and two, when $\hat{e} = \pm\hat{e}'$, which is the group A($\hat{e}$)×T. Collecting them all together, we have the eight subgroups listed in Table I.

### TABLE I: THE EIGHT SUBGROUPS

| | | |
|---|---|---|
| S×T | | S |
| A($\hat{e}$)×T | A($\hat{e}$)×M($\hat{e}'$) ($\hat{e}\cdot\hat{e}' = 0$) | A($\hat{e}$) |
| T | M($\hat{e}'$) | $\mathcal{I}\times\mathcal{I}$ |

In Table II, we have given the eight broken symmetry adopted forms of the Green function and the corresponding self-energy derived from the considerations given above. There is a corresponding set of self-energy functional that goes with these, which are also given in this Table.

# TABLE II: BROKEN SYMMETRY ADAPTED GREEN FUNCTION SOLUTIONS OF THE DYSON EQUATION

| Invariance involving Time → <br><br> Invariance involving Spin ↓ | Group T of Time Reversal | Group $M(\hat{e}')$ consisting of T and $\pi$ rotation about $\hat{e}'$-axis. | I |
|---|---|---|---|
| Group S of all spin rotations | (PARAMAGNETIC SYSTEM) <br> $G_1 = 1/2(g_n)$, with $\vec{j} = 0$. <br><br> $\Sigma_1 = \sigma_n$ <br> Non-magnetic insulator. | | Charge current wave system <br> $G_2 = 1/2(g_n)$ <br> with $\vec{j} \neq 0$. <br> $\Sigma_2 = \sigma_n$ <br> Nonmagnetic metal or semiconductor |
| Group $A(\hat{e})$ of spin rotations about $\hat{e}$-axis.(Axial) | Axial spin current wave system <br> $G_3 = 1/2$ <br> $\left(g_n + (\vec{\tau}\cdot\hat{e})g_s^{//}\right)$ <br> with $\vec{j} = 0$ and <br> $\breve{j}_S \neq 0$. <br> $\Sigma_3 = $ <br> $\left(\sigma_n + (\vec{\tau}\cdot\hat{e})\sigma_s^{//}\right)$ <br> Insulating $\hat{e}$−axis antiferromagnet | Axial spin density wave system <br> $G_4 = 1/2$ <br> $\left(g_n + (\vec{\tau}\cdot\hat{e})g_s^{//}\right)$ <br> with $\vec{j} = 0$ and <br> $\breve{j}_S = 0$. <br> $\Sigma_4 = $ <br> $\left(\sigma_n + (\vec{\tau}\cdot\hat{e})\sigma_s^{//}\right)$ <br> Insulating axial antiferromagnet. <br> $(\hat{e}'\cdot\hat{e} = 0)$ | Axial spin wave system <br> $G_5 = 1/2$ <br> $\left(g_n + (\vec{\tau}\cdot\hat{e})g_s^{//}\right)$ <br> with $\vec{j} \neq 0$ and <br> $\breve{j}_S \neq 0$. <br> $\Sigma_5 = $ <br> $\left(\sigma_n + (\vec{\tau}\cdot\hat{e})\sigma_s^{//}\right)$ <br> Itinerant electron axial antiferromagnet |
| $\mathcal{I}$ | General spin current wave system <br> $G_6 = \frac{1}{2}(g_n + \vec{\tau}\cdot\vec{g}_s)$ <br> such that <br> $\vec{j} = 0; \vec{s} = 0, but$ <br> $\breve{j}_S \neq 0$. <br> $\Sigma_6 = \left(\sigma_n + \vec{\tau}\cdot\vec{\sigma}_s\right)$ <br> Insulating spin current density state (antiferro.) | General spin density wave system <br> $G_7 = 1/2$ <br> $\begin{pmatrix} g_n + (\vec{\tau}\cdot\hat{e}')g_s^{//} + \\ (\vec{\tau} - \hat{e}(\vec{\tau}\cdot\hat{e}'))\cdot\vec{g}_s^{\perp} \end{pmatrix}$ <br> $\Sigma_7 = $ <br> $\begin{pmatrix} \sigma_n + (\vec{\tau}\cdot\hat{e}')\sigma_s^{//} + \\ (\vec{\tau} - \hat{e}(\vec{\tau}\cdot\hat{e}'))\cdot\vec{\sigma}_s^{\perp} \end{pmatrix}$ <br> Itinerant electron helical SDW state (e.g. Overhauser) | General spin wave system (with no constraint on spin density vector) <br> $G_8 = \frac{1}{2}(g_n + \vec{\tau}\cdot\vec{g}_s)$. <br> $\Sigma_8 = \left(\sigma_n + \vec{\tau}\cdot\vec{\sigma}_s\right)$ <br> Most general itinerant electron SDW state |

In this Table, we have used the invariance of the continuity equations, eqs.(25, 26) under time-reversal operation which in turn lead to respective transformation properties of particle-, particle-current-, spin-, and spin-current- densities:

$$n \to n; \quad \vec{j} \to -\vec{j}; \quad \vec{s} \to -\vec{s}; \text{ and } \vec{j}_S \to \vec{j}_S. \tag{30}$$

This representation is equivalent to but more physical than the one given by Fukutome [2], who expresses his results in terms of the real and imaginary parts of the density matrix. It may be worth pointing out that several types of itinerant spin-density-wave states occurring in rare earth systems were discussed before in a phenomenological way [18].

### III. LINEAR RESPONSE FUNCTIONS - SYMMETRY CONSIDERATIONS

The well-known theory of linear response functions in itinerant magnetic systems may be expressed in terms of the functional derivatives of the inverse Green function, eq.(15), with respect to the external fields $f^\alpha$, of eqs. (6,7). From these one examines the possible collective excitations in the system, (see for example, [13,15]). Additionally these response functions allow us to test the stability of the state (for example, in [13, 15]) that is obtained as the solution of eq.(9 or equivalently 15). Alternately, the stability question may be reformulated as the criterion of minimum free energy, as was originally done in [14] and in a more physical way recently in [16] where a localized site-representation was used. Such a version is not applicable to itinerant electron system, where one has contribution from the mobile electrons as is evident from the work of [13,15]. One way to realize this is to note that in eq.(26) there is a contribution from the divergence of the spin-current which would be absent when there are no itinerant electrons. To bring out this feature, instead of repeating the work in [13, 15], we give here a brief outline of the linear response theory. The various types of response functions, $\chi_{\alpha\beta}$, can be expressed as appropriate variational derivatives, which in turn are expressed in terms of the corresponding appropriate variational derivatives of the self-energy:

$$\chi_{\alpha\beta}(12) = -i\left\{ \left\langle T\left(\hat{s}_\alpha(1)\hat{s}_\beta(2)\right)\right\rangle - s_\alpha(1)s_\beta(2) \right\} = \frac{\delta s_\alpha(1)}{\delta f^\beta(2)} = -iTrtr\,\tau_\alpha \frac{\delta G(11^+)}{\delta f^\beta(2)} =$$

$$-iTrtr\int \tau_\alpha G \Lambda_\beta G; \quad \Lambda_\beta(12;3) = -\frac{\delta G^{-1}(12)}{\delta f^\beta(3)} = \delta(12)\delta(13)\tau_\beta + \frac{\delta \Sigma(12)}{\delta f^\beta(3)}.$$

Recalling the definitions given in eqs.(7, 20, 21), we have 16 response functions, involving the particle density and the three components of the spin vector density. In the localized electron scheme, the cross terms involving the density and spin vector will not appear. This is another important difference between the localized and itinerant electron systems. From such a linear response theory, one often deduces the low-energy collective excitations in the system, such as spin waves in ferro-, antiferro- and SDW- systems. Such a discussion may be found, for example, in [15].

The objective in [16] was to derive expressions for the energy of the known localized effective spin models of magnetic systems. From this one can also deduce the low energy collective excitations in the system. In the itinerant electron system, it should be pointed out that there are important contributions not found in the localized spin systems, as is evident from [15], for example. We adopt the alternate formulation in this section to obtain a generalization of the "local force theorem" in [16] which is applicable to both the localized and the itinerant systems.

Unlike in the general linear response theory described above, we now consider the effect of an infinitesimal rotation of the spin about a general direction denoted by a unit vector $\hat{e}$, $\delta\vec{\theta}^{\perp} = \hat{e}\delta\theta$, obtained from eq.(28): $\delta U = (1 - i\delta\vec{\theta}^{\perp} \cdot \vec{\tau}/2)$, on the free energy, eq.(16), in the same fashion as in [16] holding G fixed:

$$\delta\Omega = -Tr\, tr\, G(\delta\Sigma);$$

$$\delta\Sigma = i\frac{\delta\vec{\theta}}{2} \cdot (\vec{\tau}\Sigma - \Sigma\vec{\tau}) = i\frac{\delta\vec{\theta}}{2} \cdot (\vec{\tau}(\vec{\tau}\cdot\vec{\sigma}_S) - (\vec{\tau}\cdot\vec{\sigma}_S)\vec{\tau})$$

$$= -\vec{\tau}\cdot(\delta\vec{\theta} \times \vec{\sigma}_S),\, and\, so$$

$$\delta\Omega = \delta\vec{\theta} \cdot tr(\vec{\sigma}_S \times \vec{g}_S) \equiv \delta\vec{\theta} \cdot \vec{V}.$$

(31)

Here we used the representations given in eqs.(18, 19) and some well-known identities to arrive at the expression for the torque, $\vec{V}^{\perp}$, due to the rotation. This expression differs from the one given in [16] because of a factor of half in our definition of the spinor-Green function in contrast to theirs. As in [16], the Dyson equation gives us the sum rules on the components of the spinor Green function, after using eq.(15):

$$GG^{-1} = 1 = G(G_0^{-1} - \Sigma) = \frac{1}{2}(g_n + \vec{\tau}\cdot\vec{g}_S)(G_0^{-1} - \sigma_n - \vec{\tau}\cdot\vec{\sigma}_S);$$

$$G^{-1}G = 1 = (G_0^{-1} - \Sigma)G = (G_0^{-1} - \sigma_n - \vec{\tau}\cdot\vec{\sigma}_S)\frac{1}{2}(g_n + \vec{\tau}\cdot\vec{g}_S).$$

(32)

From which we get the relations:

$$(G_0^{-1} - \sigma_n)g_n - \vec{\sigma}_S \cdot \vec{g}_S = 2.$$

$$(G_0^{-1} - \sigma_n)\vec{g}_S - \vec{\sigma}_S g_n - i\vec{\sigma}_S \times \vec{g}_S = 0.$$

(33a,b)

$$g_n(G_0^{-1} - \sigma_n) - \vec{g}_S \cdot \vec{\sigma}_S = 2.$$

$$\vec{g}_S(G_0^{-1} - \sigma_n) - g_n\vec{\sigma}_S - i\vec{g}_S \times \vec{\sigma}_S = 0.$$

(34a,b)

From these we derive the following sum rules corresponding to those given in [16] but in a coordinate representation valid for both itinerant and localized electron systems. Multiplying (33a) by $g_n$ on the left and (34a) on the right we obtain

$$2g_n = g_n(G_0^{-1} - \sigma_n)g_n - g_n(\vec{\sigma}_S \cdot \vec{g}_S)$$
$$= g_n(G_0^{-1} - \sigma_n)g_n - (\vec{g}_S \cdot \vec{\sigma}_S)g_n. \tag{35}$$

Multiplying (33a) by $\vec{g}_S$ on the left and using (34b) we obtain
$$2\vec{g}_S = g_n\vec{\sigma}_S g_n - \vec{g}_S(\vec{\sigma}_S \cdot \vec{g}_S) + i(\vec{g}_S \times \vec{\sigma}_S)g_n$$
$$= g_n\vec{\sigma}_S g_n - (\vec{g}_S \cdot \vec{\sigma}_S)\vec{g}_S + ig_n(\vec{\sigma}_S \times \vec{g}_S). \tag{36}$$

The second expression is obtained by multiplying (34a) by $\vec{g}_S$ on the right and using (33b). From eqs.(31, 36), we obtain the general expression for the torque vector

$$\vec{V} = tr(\vec{\sigma}_S \times \vec{g}_S) = -\frac{1}{2}tr\left\{\vec{\sigma}_S \times \left(-g_n\vec{\sigma}_S g_n + (\vec{g}_S \cdot \vec{\sigma}_S \cdot)\vec{g}_S - ig_n(\vec{\sigma}_S \times \vec{g}_S)\right)\right\} \tag{37}$$

By integrating back the expression given by eq.(31), effectively the spin only part of the free energy may be rewritten as
$$\Omega_{sp} \cong trTr(G\Sigma) \cong tr(\vec{g}_S \cdot \vec{\sigma}_S)$$
$$= -\frac{1}{2}tr\left\{(\vec{g}_S \cdot \vec{\sigma}_S)(\vec{g}_S \cdot \vec{\sigma}_S) - (\vec{\sigma}_S g_n) \cdot (\vec{\sigma}_S g_n) - i((\vec{\sigma}_S \times g_n\vec{\sigma}_S) \cdot \vec{g}_S)\right\}. \tag{38}$$

Here eq.(36) was used in further simplification. The second approximation symbol is because we have dropped the spin-independent contribution, $trg_n\sigma_n$, arising in the first expression. Note however, that from eq.(35), $g_n, \sigma_n$ depend on the spin vector. Had we kept this contribution, we would have additional contributions due to particle density-density and particle density and spin density vector mentioned in the linear response theory at the beginning of this Section. For completeness, we give this here: $\Omega_{ns} \cong trg_n\sigma_n \cong \frac{1}{2}tr\left\{\sigma_n g_n(G_0^{-1} - \sigma_n)g_n - \sigma_n g_n(\vec{\sigma}_S \cdot \vec{g}_S)\right\}$. Here eq.(35) was used in further simplification. It should be noted that the notation tr used here is as defined before in eq.(16a). In the site-local representation used in [16], eq.(38) goes over to that given there.

Using Table II, we have eight types of Green functions and their corresponding self-energies associated with the allowed broken symmetry solutions. Using these in the general expressions for the spin-spin interaction energies obtained above, one can deduce the structures of the corresponding spin-spin contribution to the free energy of the system.

### IV. IMPLICATION TO VECTOR - SPIN - DENSITY - FUNCTIONAL THEORY

The development of the vector-spin-density-functional theory especially the local (LSD) approximation, has in recent years produced a much better theoretical understanding of itinerant magnetism. (See for example [5, 9, 10].) In particular, in [15], a spinor-Green function version of the LSD theory was developed where the self-energy in eq.(9 or 15) is taken to be of the form, local in space and time,

$$\Sigma_{LSD}(12) \cong \{V_{xc}[n,\vec{s};\vec{r_1}] + \vec{\tau}\cdot\overset{\leftrightarrow}{W}_{xc}[n,\vec{s};\vec{r_1}]\}\delta(12), \qquad (39)$$

where $V_{xc}$ is the spin-scalar part of the self-energy, which is in general a local functional of particle density, n, and spin density vector, $\vec{s}$, whereas $\overset{\leftrightarrow}{W}_{xc}$ is its spin-vector counterpart, for describing the itinerant magnetic systems. The approximate forms for these functionals arise as the respective functional derivatives of the exchange-correlation energy, $E_{xc}[n,\vec{s}] \equiv E_{xc}[s_\alpha]$, (using the notation of eq.(7)) of a spin-polarized homogeneous electron gas (see references in [5], for example). In fact, $V_{xc} = \delta E_{xc}/\delta n$, $\overset{\leftrightarrow}{W}_{xc} = \delta E_{xc}/\delta \vec{s}$, or more generically written as $V_{xc}^{(\alpha)} = \delta E_{xc}/\delta s_\alpha$, so that $\Sigma_{LSD} = \tau_\alpha V_{xc}^{(\alpha)}$ The resulting equation is the LSD equation,

$$\left[i\frac{\partial}{\partial t_1} + \frac{\vec{\nabla}_1^2}{2m} - V_C(1) - w_n(1) - \vec{\tau}\cdot\vec{f}(1)\right]G(12) -$$

$$- \{V_{xc}[n,\vec{s};1] + \vec{\tau}\cdot\overset{\leftrightarrow}{W}_{xc}[n,\vec{s};1]\}G(12) = \delta(12). \qquad (40)$$

This is solved self-consistently by numerical methods.

The continuity equations for the particle- and spin-density derived from this LSD eq.(40) now take the form

$$\frac{\partial n(1)}{\partial t_1} + \vec{\nabla}_1 \cdot \vec{j}(1) = 0;$$

$$\frac{\partial \vec{s}(1)}{\partial t_1} + \vec{\nabla}_1 \cdot \overset{\leftrightarrow}{j}_S(1) = \overset{\leftrightarrow}{W}_{xc} \times \vec{s}(1). \qquad (41)$$

The second equation for the spin-density vector has a contribution from the torque due to the spin polarization of the spin system, besides having a contribution from the divergence of the moving spins producing a spin-current density.

Proceeding as in eqs.(33, 34), we have the equations

$$\left(G_0^{-1} - V_{xc}\right)g_n - \left(\overset{\leftrightarrow}{W}_{xc}\right)\cdot\vec{g}_S = 2.$$

$$\left(G_0^{-1} - V_{xc}\right)\vec{g}_S - \left(\overset{\leftrightarrow}{W}_{xc}\right)g_n - i\left(\overset{\leftrightarrow}{W}_{xc}\right)\times\vec{g}_S = 0. \qquad (42a, b)$$

$$g_n\left(G_0^{-1} - V_{xc}\right) - \vec{g}_S\cdot\left(\overset{\leftrightarrow}{W}_{xc}\right) = 2.$$

$$\vec{g}_S\left(G_0^{-1} - V_{xc}\right) - g_n\left(\overset{\leftrightarrow}{W}_{xc}\right) - i\vec{g}_S\times\left(\overset{\leftrightarrow}{W}_{xc}\right) = 0. \qquad (43a,b)$$

From these we obtain the expressions

$$2g_n = g_n\left(G_0^{-1} - V_{xc}\right)g_n - g_n\left(\overset{\leftrightarrow}{W}_{xc}\cdot\vec{g}_S\right)$$

$$= g_n\left(G_0^{-1} - V_{xc}\right)g_n - \left(\vec{g}_S\cdot\overset{\leftrightarrow}{W}_{xc}\right)g_n. \qquad (44a,b)$$

$$2\vec{g}_S = g_n(\vec{W}_{xc}^\perp)g_n + i(\vec{g}_S \times \vec{W}_{xc}^\perp)g_n - \vec{g}_S(\vec{W}_{xc}^\perp \cdot \vec{g}_S),$$
$$= g_n(\vec{W}_{xc}^\perp)g_n + ig_n(\vec{W}_{xc} \times \vec{g}_S) - (\vec{g}_S \cdot \vec{W}_{xc})\vec{g}_S.$$
(45a,b)

Finally the expression for the spin-only contribution to the free energy in the LSD case is obtained in a manner similar to the one given in Sec.III, eq.(38):

$$\Omega_{sp}^{LSD} = -\frac{1}{2}tr\left\{(\vec{g}_S \cdot \vec{W}_{xc})(\vec{g}_S \cdot \vec{W}_{xc}) - (g_n\vec{W}_{xc}) \cdot (g_n\vec{W}_{xc}) - ig_n(\vec{W}_{xc} \times \vec{g}_S) \cdot \vec{W}_{xc}\right\}.$$
(46)

The corresponding particle density and spin density contribution to free energy is

$$\Omega_{ns}^{LSD} \cong trg_nV_{xc} = \frac{1}{2}tr\left\{V_{xc}g_n(G_0^{-1} - V_{xc})g_n - V_{xc}g_n(\vec{W}_{xc} \cdot \vec{g}_S)\right\}.$$ In eq.(46), as in eq.(38), the first term represents a multi-spin interaction energy involving four spins or more, the middle term is like the spin-spin interaction involving two spins or more, whereas the last term is a Dzialoshinskii-Moriya - type interaction, involving three or more spins. In [15], a spatially slowly varying approximation was considered in an approximate way and only the middle term resembling an effective Heisenberg spin-spin interaction energy was derived along with the cross terms containing particle density and spin density. The higher order spin terms and the Dzialoshinskii-Moriya terms did not appear in that derivation because the self-consistent relations as in eqs.(44, 45) were not invoked. The above remark arises from the observation that the vector part of the self-energy functional, $\vec{W}_{xc}$, is an odd functional of the spin vector beginning with a linear functional of the spin vector.

In Table III given below, the eight types of Green functions and the corresponding self-energies are given associated with the various broken symmetry types for the case of LSD theory. From the above general expressions for the free energy, one may then deduce the structure spin dependent energies that follow for each of these cases. The phenomenological description given [18] is here supported from the considerations of group theory.

Applying the linear response theory outlined in Sec.III to the LSD scheme, one obtains a tensor: $\frac{\delta\Sigma}{\delta f^\beta} = tr\frac{\delta\Sigma}{\delta s^\gamma}\frac{\delta s_\gamma}{\delta f^\beta}(chain\ rule) = \tau_\alpha tr\frac{\delta^2 E_{xc}}{\delta s^\alpha \delta s^\gamma}\frac{\delta s_\gamma}{\delta f^\beta}$, as was shown in [15]. The broken symmetry considerations leading to Table III may be applied to this tensor to deduce the corresponding eight structures. As shown in [15] and more recently in [10], the use of homogeneous electron gas results in LSD requires a subtle and important modification in incorporating the vector nature of the spin density, when studying the spin wave properties of itinerant magnets. Another way of expressing this point is that the traditional electron gas theory leads to Ising-like treatment of the spins which is converted into a Heisenberg-like treatment (see for example, [9]) by a spin rotation. This does not lead to correct answers as was shown in [10] and this is due to the subtle nature of the treatment of the spin vector in the theory. In [10], a perturbation theory approach has been presented to include this feature, thus making a significant difference.

# TABLE III: BROKEN SYMMETRY ADAPTED GREEN FUNCTION SOLUTIONS OF THE LSD EQUATION

| Invariance involving Time → <br><br> Invariance involving Spin ↓ | Group T of Time Reversal | Group $M(\hat{e}')$ consisting of T and $\pi$ rotation about $\hat{e}'$-axis. | I |
|---|---|---|---|
| Group S of all spin rotations | (PARAMAGNETIC SYSTEM) <br> $G_1 = 1/2(g_n)$, with $\vec{j} = 0$. <br><br> $\Sigma_1 = \delta(12) V_{xc}$ <br> Non-magnetic insulator. | | Charge current wave system <br> $G_2 = 1/2(g_n)$ <br> with $\vec{j} \neq 0$. <br> $\Sigma_2 = \delta(12) V_{xc}$ <br> Nonmagnetic metal or semiconductor |
| Group $A(\hat{e})$ of spin rotations about $\hat{e}$-axis.(Axial) | Axial spin current wave system <br> $G_3 = 1/2$ <br> $\left(g_n + (\vec{\tau}\cdot\hat{e})g_s^{//}\right)$ <br> with $\vec{j} = 0$ and $\breve{j}_S \neq 0$. <br> $\Sigma_3 = \delta(12)$ <br> $\left(V_{xc} + (\vec{\tau}\cdot\hat{e})\vec{W}_{xc}^{//}\right)$ <br> Insulating $\hat{e}$-axis antiferromagnet | Axial spin density wave system <br> $G_4 = 1/2$ <br> $\left(g_n + (\vec{\tau}\cdot\hat{e})g_s^{//}\right)$ <br> with $\vec{j} = 0$ and $\breve{j}_S = 0$. <br> $\Sigma_4 = \delta(12)$ <br> $\left(V_{xc} + (\vec{\tau}\cdot\hat{e})\vec{W}_{xc}^{//}\right)$ <br> Insulating axial antiferromagnet. <br> $(\hat{e}'\cdot\hat{e} = 0)$ | Axial spin wave system <br> $G_5 = 1/2$ <br> $\left(g_n + (\vec{\tau}\cdot\hat{e})g_s^{//}\right)$ <br> with $\vec{j} \neq 0$ and $\breve{j}_S \neq 0$. <br> $\Sigma_5 = \delta(12)$ <br> $\left(V_{xc} + (\vec{\tau}\cdot\hat{e})\vec{W}_{xc}^{//}\right)$ <br> Itinerant electron axial antiferromagnet |
| I | General spin current wave system <br> $G_6 = \frac{1}{2}(g_n + \vec{\tau}\cdot\vec{g}_s)$ <br> such that $\vec{j} = 0; \vec{s} = 0, but$ <br> $\breve{j}_S \neq 0$. <br> $\Sigma_6 = \delta(12)$ <br> $\left(V_{xc} + \vec{\tau}\cdot\vec{W}_{xc}\right)$ <br> Insulating spin current density state (antiferro.) | General spin density wave system <br> $G_7 = 1/2$ <br> $\begin{pmatrix} g_n + (\vec{\tau}\cdot\hat{e}')g_s^{//} + \\ (\vec{\tau} - \hat{e}(\vec{\tau}\cdot\hat{e}'))\cdot\vec{g}_s^{\perp} \end{pmatrix}$ <br> $\Sigma_7 = \delta(12)$ <br> $\begin{pmatrix} V_{xc} + (\vec{\tau}\cdot\hat{e}')\vec{W}_{xc}^{//} + \\ (\vec{\tau} - \hat{e}(\vec{\tau}\cdot\hat{e}'))\cdot\vec{W}_{xc}^{\perp} \end{pmatrix}$ <br> Itinerant electron helical SDW state (e.g. Overhauser) | General spin wave system (with no constraint on spin density vector) <br> $G_8 = \frac{1}{2}(g_n + \vec{\tau}\cdot\vec{g}_s)$. <br> $\Sigma_8 = \delta(12)$ <br> $\left(V_{xc} + \vec{\tau}\cdot\vec{W}_{xc}\right)$ <br> Most general itinerant electron SDW state |

# V. SUMMARY AND CONCLUDING REMARKS

In summary, the structures of the Green functions and their associated self energy arising from group theoretical considerations of the spin rotation and time-reversal invariance are given in Table II for general magnetic many-electron systems and in Table III, for the vector-spin-density functional formalism of the itinerant electron systems. This is expected to systematize the procedure in the analysis of magnetic structures that may appear in magnetic nanometric systems and in magnetic atomic clusters, just as the earlier similar work of Fukutome did systematize the Hartree-Fock solutions of magnetic states of molecular systems. We also consider consequences of this by setting up the linear response theory and an alternate version of it in the form of "effective spin Hamiltonian", to exhibit the differences between localized electron systems with those where the electrons are itinerant. We hope that the search for various types of magnetic structures in nanometric [8] and atomic cluster [11] systems, particularly when the clusters are deposited on semiconductor substrates [12], will be systematized by the procedure given here. In these systems, as seen from the work in [11], magnetic properties are correlated with the geometric structures of the clusters; the group theoretical analysis presented here is expected to make the search for this feature systematic. In this context, the work presented here incorporated into that in [17], may be expected to lead to efficient procedures for computation of magnetic properties in nanometric systems and in atomic clusters.

We have here pointed out the significance of incorporating the vector nature of spin density in LSD theory, particularly in the modification needed in the traditional use of the homogeneous electron gas results. In this context, the recent work in [10] should be mentioned as an important step in a proper treatment of the vector spin. In [10], the transverse part of the vector-spin was incorporated in a perturbative way and was shown to lead to a better understanding of magnetism of Iron than previously [9]. The localized treatment of spin interactions deduced in [16] is here generalized to itinerant electron magnetic systems and the differences arising from this are spelled out. It may also be pointed out that the phenomenology of the various types of SDW structures given in [18] may be deduced from Table III.

In this paper, we have not included the gauge group needed to incorporate superconducting phases nor have we included the lattice translation group. The addition of the translation group into the considerations given here brings in the irreducible representation characterized by the q-vector associated with the Brillouin zone and the little group of the q-vector. The gauge group is also useful particularly because of the newly discovered high $T_C$ superconductors involving d- and s- wave pairing possess interesting vortex structures. Inclusion of these within the mean field (HF) theory has been reported in some special cases [2, 4]. The generalization of all these features in the Green function framework will be addressed in a future communication.

## ACKNOWLEDGMENTS

AKR is supported in part by the Office of Naval Research. MM acknowledges the financial support of ASEE Summer Faculty Program which enabled him to continue collaboration at the Naval Research Laboratory, Washington, D. C.